%% file: main.tex
\documentclass[sigconf]{acmart}

\usepackage{latexsym}
\usepackage[utf8]{inputenc}
\usepackage{todonotes} 
\usepackage{hyperref}
\usepackage{graphicx}
\usepackage{float}
\usepackage{url}
\usepackage{amsmath}
\usepackage{bm}
\usepackage{amsthm}
\usepackage{multirow}
\usepackage{booktabs}
\usepackage{arydshln}
\usepackage{xspace}
\usepackage{subcaption}
\usepackage{scalerel}
\usepackage{mathtools}
\usepackage{soul}
\usepackage{color}

\input{dlbook_mathcommands}

\copyrightyear{2020} 
\acmYear{2020} 
\setcopyright{acmcopyright}\acmConference[SIGIR '20]{Proceedings of the 43rd International ACM SIGIR Conference on Research and Development in Information Retrieval}{July 25--30, 2020}{Virtual Event, China}
\acmBooktitle{Proceedings of the 43rd International ACM SIGIR Conference on Research and Development in Information Retrieval (SIGIR '20), July 25--30, 2020, Virtual Event, China}
\acmPrice{15.00}
\acmDOI{10.1145/3397271.3401280}
\acmISBN{978-1-4503-8016-4/20/07}

\newcommand{\matchpyramid}{\text{MatchPyramid}\xspace}
\newcommand{\pacrr}{\text{PACRR}\xspace}
\newcommand{\knrm}{\text{KNRM}\xspace}
\newcommand{\convknrm}{\text{ConvKNRM}\xspace}
\newcommand{\bertbase}{\text{BERT-Base}\xspace}
\newcommand{\bertlarge}{\text{BERT-Large}\xspace}

\newcommand{\matchpyramidrnd}{$\text{MatchPyramid}_{RND}$\xspace}
\newcommand{\pacrrrnd}{$\text{PACRR}_{RND}$\xspace}
\newcommand{\knrmrnd}{$\text{KNRM}_{RND}$\xspace}
\newcommand{\convknrmrnd}{$\text{ConvKNRM}_{RND}$\xspace}

\newcommand{\docfactor}{mag\xspace}
\newcommand{\docfactortc}{\texttt{TF}\xspace}
\newcommand{\docfactorbool}{\texttt{Boolean}\xspace}

\newcommand{\qbiasrab}{\text{qRaB}\xspace}
\newcommand{\qbiasArab}{\text{qARaB}\xspace}

\newcommand{\biasrab}{\text{RaB}\xspace}
\newcommand{\biasArab}{\text{ARaB}\xspace}

\newif\ifworkinprogress
\workinprogresstrue

\ifworkinprogress
  \newcommand{\ms}[1]{\textcolor{blue}{[Markus] #1}}
  \newcommand{\nr}[1]{\textcolor{darkgreen}{[Navid] #1}}
\else
  \newcommand{\ms}[1]{}
  \newcommand{\nr}[1]{}
\fi

\begin{document}

\fancyhead{}

\title{Do Neural Ranking Models Intensify Gender Bias?}

\author{Navid Rekabsaz}
\email{navid.rekabsaz@jku.at}
\affiliation{%
  \institution{Johannes Kepler University}
}

\author{Markus Schedl}
\email{markus.schedl@jku.at}
\affiliation{%
  \institution{Johannes Kepler University}
}

\renewcommand{\shortauthors}{Rekabsaz and Schedl}

\begin{abstract}
\vspace{-0.1cm}
Concerns regarding the footprint of societal biases in information retrieval (IR) systems have been raised in several previous studies. In this work, we examine various recent IR models from the perspective of the degree of gender bias in their retrieval results. To this end, we first provide a bias measurement framework which includes two metrics to quantify the degree of the unbalanced presence of gender-related concepts in a given IR model's ranking list. To examine IR models by means of the framework, we create a dataset of non-gendered queries, selected by human annotators. Applying these queries to the MS~MARCO Passage retrieval collection, we then measure the gender bias of a BM25 model and several recent neural ranking models. The results show that while all models are strongly biased toward male, the neural models, and in particular the ones based on contextualized embedding models, significantly intensify gender bias. Our experiments also show an overall increase in the gender bias of neural models when they exploit transfer learning, namely when they use (already biased) pre-trained embeddings.\footnote{In this work, we treat gender as a binary construct. We acknowledge that this choice neglects the broad meaning of gender, but the decision is necessary for practical reasons given the technical constraints and the limited scope of the work.} 
\vspace{-0.1cm}

\end{abstract}



\maketitle

\vspace{-0.2cm}
\section{Introduction}
\vspace{-0.1cm}
\label{sec:introduction}
\input{1-introduction}

\vspace{-0.2cm}
\section{Retrieval Bias Measurement Framework}
\label{sec:framework}
\input{2-framework}

\vspace{-0.2cm}
\section{Experiment Design}
\label{sec:experiments}
\input{3-experiments}

\vspace{-0.2cm}
\section{Results}
\vspace{-0.1cm}
\label{sec:results}
\input{4-results}

\vspace{-0.5cm}
\section{Conclusion and Future Work}
\label{sec:conclusion}
\vspace{-0.1cm}
This work takes a first step in measuring the degree of gender bias in retrieval models and studying it in neural IR models. We propose a novel framework to measure gender bias in retrieval results and provide a set of human-annotated non-gendered queries. By submitting these queries to various IR models, we show that the studied neural ranking models intensify gender bias towards male concepts in comparison with BM25. The fine-tuned BERT models show the highest degrees of bias. We also observe that the neural ranking models (excluding the BERT ones) generally increase gender bias when they use transfer learning.  

Future research following this work further investigates the relation between bias and relevance in retrieval. Based on the results of this study, the gender bias values of the neural IR models do not fully correlate with their performance. This encourages a deeper analysis of the neural models. Another direction of research is the study of methods to eliminate gender bias in neural IR models while preserving their effectiveness, especially in the light of literature on fairness in ranking and embedding debiasing. Finally, exploring other metrics of bias measurement, as well as studying the relations between the metrics and the human perception of bias in retrieval results, are other future avenues of this work.

\vspace{-0.3cm}
\begin{acks}
\vspace{-0.1cm}
Many thanks to Sophia Freynhofer for her help and advice on designing crowd sourcing experiments. 
\end{acks}

\vspace{-0.3cm}
\bibliographystyle{ACM-Reference-Format}
\bibliography{references}

\end{document}

%% file: dlbook_mathcommands.tex
\def\eqref#1{equation~\ref{#1}}

\def\1{\bm{1}}

\DeclareMathAlphabet{\mathsfit}{\encodingdefault}{\sfdefault}{m}{sl}
\SetMathAlphabet{\mathsfit}{bold}{\encodingdefault}{\sfdefault}{bx}{n}

\def\sG{{\mathbb{G}}}



%% file: 1-introduction.tex
The existence of various types of bias in IR, and the concerns regarding its implications on IR applications, have been pointed out in several previous studies~\cite{baeza2018bias,kay2015unequal,chen2018investigating,biega2018equity}. As an example of gender bias in IR, conducting an ad-hoc retrieval of the query \emph{nurse} -- a gender neutral word -- using a baseline exact matching IR model (such as BM25~\cite{robertson2009probabilistic}) results in a highly disproportional presence of female-related concepts in the retrieved documents/images~\cite{kay2015unequal,otterbacher2017competent,otterbacher2018investigating}. Biased results can lead to an unfair distribution of opportunities and resources~\cite{chen2018investigating,biega2018equity}, and ``echo chamber'' effects. The problem surfaced by this example mainly originates from the intrinsic bias in collections -- documents containing the term \emph{nurse} most probably also contain female-related concepts -- and from the way various IR models estimate relevance. Our focus is on the latter, measuring gender bias in retrieval results, and comparing different IR models.


Recent advancements in neural ranking models along with the availability of collections with large amounts of relevance information have brought about remarkable improvements to retrieval performance. The recent neural ranking models go beyond exact matching of terms by exploiting the semantic space of embedding vectors~\cite{dai2018convolutional,xiong2017end,hui2017pacrr,hofstatter2019effect,nogueira2019passage,pang2016text}. They also typically benefit from transfer learning, i.e.\ initializing a set of model parameters with pre-trained word embeddings such as GloVe~\cite{pennington2014glove}, or contextualized embeddings such as BERT~\cite{devlin2019bert}. 

However, as shown previously~\cite{elazar2018adversarial}, the embeddings in neural networks models, while being capable of capturing effective semantic relations, may also encode societal biases present in training data. It is also well-known that the pre-trained embeddings, whether word or contextualized ones, reflect societal biases which originate from their underlying corpora~\cite{caliskan2017semantics,bolukbasi2016man,zhao2019gender}.\footnote{For instance, some gender-neutral words such as \emph{housekeeper} and \emph{nurse} are strongly associated in these embeddings with female-related words such as \emph{she} and \emph{woman}.} Unlike the previous studies, this work investigates these aspects in IR, by studying the degree of gender bias in neural ranking models, and the effect of transfer learning on them.

This study delivers three contributions. First, we provide a framework to measure gender bias of retrieval models. To this end, we first use a set of highly gender-related words to measure \emph{document female/male magnitude}, namely the degree of female-/male-related concepts in a document. Next, following previous work~\cite{kulshrestha2017quantifying,yang2017measuring}, we define two retrieval gender bias metrics. These metrics calculate the differences of the female/male magnitudes of the documents retrieved by a given model, averaged over a set of given queries. 


To enable testing various IR models, in the next contribution, we provide a set of \emph{non-gendered queries}, i.e.\ information needs that do not contain any element referring to a specific gender. We conduct human annotation experiments to identify the set of non-gendered queries among the queries of the development set of the MS~MARCO Passage Retrieval collection~\cite{nguyen2016ms}. 

Using the framework and the non-gendered queries, in our third contribution, we measure the gender bias of a BM25 model, and six neural ranking models, including two fine-tuned BERT models. We conduct our experiments on the MS~MARCO Passage Retrieval collection. For the neural ranking models (except the BERT-based ones), we experiment with both random initialization as well as transferring a pre-trained GloVe embedding. 




The results reveal the existence of significant amounts of gender bias (towards male) in the retrieved documents of all the models. The neural ranking models demonstrate considerably higher degrees of bias in comparison with BM25 on both retrieval bias metrics and various ranking cutoffs. In particular, the fine-tuned BERT models, despite achieving the best retrieval performance, show the overall highest gender bias. Finally, comparing the models with and without pre-trained word embeddings, we observe that applying transfer learning intensifies gender bias in neural ranking models.


The remainder of the paper is organized as follows: We describe our retrieval bias measurement framework in Section~\ref{sec:framework}. The experiment design and the created dataset are described in Section~\ref{sec:experiments}. We present and discuss the results in Section~\ref{sec:results}.


%% file: 2-framework.tex
Our framework for measuring gender bias in retrieval models first proposes an approach to quantify female/male magnitude of a document, followed by suggesting two metrics to measure the overall degree of gender bias in the ranking lists of a model, given a set of queries. We explain the details of the framework in the following.

\vspace{-0.3cm}
\subsection{Document Gender Magnitude Measurements}
To measure the degree of presence of female/male concepts in a document -- gender magnitude -- we first define gender concepts using a set of highly representative gender words, referred to as \emph{gender definitional words}. This approach is commonly used in previous studies to define gender concepts and to measure gender bias, especially in embedding models~\cite{caliskan2017semantics,bolukbasi2016man}. Such a set typically contains words such as \emph{she}, \emph{woman}, \emph{her} for female, and \emph{he}, \emph{man}, \emph{him} for male. 

We should note that this approach is not comprehensive, namely it does not capture all the existing gender-related concepts, but approximates these using a subset of highly representative words.
Next, using the gender definitional words, we measure the female/male magnitude of a document in two variants: calculating the sum of the logarithm of the number of occurrences of the words in the document (\docfactortc), and whether any of the words exists in the document (\docfactorbool). The female magnitudes of a document in these variants are formulated as follows:
\begin{equation}
\vspace{-0.1cm}
\begin{gathered}
\textbf{\docfactortc}:\quad\docfactor^{\,f}\!(d)=\sum_{w \in \sG_{f}}{\log\#\langle w,d\rangle}\\
\textbf{\docfactorbool}:\quad\docfactor^{\,f}\!(d)=\begin{cases}
    1,& \text{if } \sum_{w \in \sG_{f}}{\#\langle w,d\rangle}> 0\\
    0,              & \text{otherwise}
\end{cases}
\label{eq:framework:docfactor}
\end{gathered}
\vspace{-0.1cm}
\end{equation}
where $\docfactor^{\,f}\!(d)$ denotes the female magnitude of the document $d$ (either using the \docfactortc or \docfactorbool variant), $\sG_{f}$ is the set of female definitional words, and $\#\langle w,d\rangle$ refers to the number of occurrences of the word $w$ in $d$. Following the same formulations, the document male-magnitude $\docfactor^{\,m}\!(d)$ is defined using the set of male definitional words $\sG_{m}$.


\vspace{-0.2cm}
\subsection{Retrieval Gender Bias Metrics}
Using the described document gender magnitude measurements, we suggest two retrieval gender bias metrics. The metrics define the gender bias of a retrieval model using the ranking lists produced by the model for a set of queries $Q$. The metrics are calculated on a cutoff $t$. We refer to the document at the $i$\textsuperscript{th} position of the ranking list, retrieved for the query $q \in Q$, as $d_i^{(q)}$.

The first metric referred to as \emph{Rank Bias (\biasrab)} is based on the averages of the gender magnitudes of the documents at the top of the ranking list. To measure retrieval bias using \biasrab, we first calculate the mean of the gender magnitudes of the top $t$ retrieved documents for the query $q$, formulated as follows for female:
\begin{equation}
\vspace{-0.1cm}
\qbiasrab^{\,f}_{t}\!(q)=\frac{1}{t}\sum_{i=1}^{t}{\docfactor^{\,f}\!\left(d_i^{(q)}\right)}
\label{eq:framework:queryfactorrab}
\vspace{-0.1cm}
\end{equation}

Using these values, the \biasrab metric of the query $q$, $\biasrab_{t}(q)$, and the \biasrab metric of the retrieval model over all the queries, $\biasrab_{t}$, are defined as follows:
\begin{equation}
\vspace{-0.1cm}
\begin{gathered}
\biasrab_{t}\!(q) = \qbiasrab^{\,m}_{t}\!(q) - \qbiasrab^{\,f}_{t}\!(q)\\ \biasrab_{t} = \frac{1}{\left|Q\right|} \sum_{q \in Q}{\biasrab_{t}\!(q)} 
\label{eq:framework:factorrab}
\end{gathered}
\vspace{-0.1cm}
\end{equation}

The second retrieval gender bias metric, originally proposed by Kulshrestha et al.~\cite{kulshrestha2017quantifying}, factors the ranking positions of documents into the \biasrab metric, following a similar approach as in the Average Precision metric. We refer to this second metric as \emph{Average Rank Bias~(\biasArab)}. The \biasArab metric first calculates the average of the \qbiasrab scores after each ranking position for each query $q$. This calculation for female is formulated as follows:
\begin{equation}
\vspace{-0.1cm}
\qbiasArab^{\,f}_{t}\!(q)=\frac{1}{t}\sum_{x=1}^{t}{\qbiasrab^{\,f}_{x}\!(q)}
\label{eq:framework:queryfactorArab}
\vspace{-0.1cm}
\end{equation}

Next, similar to \biasrab, \biasArab defines the bias of a query, $\biasArab_{t}\!(q)$, as the differences between the values of the genders, and the bias of a retrieval model is the average over all queries:
\begin{equation}
\vspace{-0.1cm}
\begin{gathered}
\biasArab_{t}\!(q) = \qbiasArab^{\,m}_{t}\!(q) - \qbiasArab^{\,f}_{t}\!(q)\\
\biasArab_{t} = \frac{1}{\left|Q\right|} \sum_{q \in Q}{\biasArab_{t}\!(q)}
\end{gathered}
\label{eq:framework:factorArab}
\vspace{-0.1cm}
\end{equation}

Both metrics, whether in query- or model-level, treat gender bias as a signed value, i.e.\ a positive value indicates bias towards male, and a negative value towards female. 

We should note two limitations of the proposed framework. First, the proposed metrics do not account for the background bias in collection, namely the uneven distribution of female versus male concepts in the collection documents. This however does not affect our experiments since our objective is to compare across IR models, where bias values are measured on the same collection. Second, the retrieval metrics are defined as the average of per-query bias values, and do not take into account the distribution of these values. Considering averaging as a first approach, we leave other metrics to define model-level bias for future directions.

%% file: 3-experiments.tex
\begin{table}
\begin{center}
\caption{The number of parameters and evaluation results of the models. GloVe consists of 109,481,400 parameters.} 
\vspace{-0.2cm}
\scriptsize
\begin{tabular}{l r r || l l}%
\toprule
\multirow{2}{*}{Ranking Model} & \multicolumn{2}{c||}{Model Parameters}  & \multicolumn{2}{c}{Evaluation} \\\cmidrule(lr){2-3} \cmidrule(lr){4-5}
 & \multicolumn{1}{c}{All} & \multicolumn{1}{c||}{Transferred} & MRR & Recall  \\\midrule

BM25 & \multicolumn{2}{c||}{--} & 0.192 & 0.398\\\hdashline
\knrmrnd & \multirow{2}{*}{109,481,411} & none & 0.213 & 0.390\\
\knrm & & GloVe & 0.230 & 0.439\\\hdashline
\matchpyramidrnd & \multirow{2}{*}{109,539,960} & none & 0.232 & 0.424\\
\matchpyramid & & GloVe & 0.240 & 0.445\\\hdashline
\pacrrrnd & \multirow{2}{*}{109,875,938} & none & 0.228 & 0.426\\
\pacrr & & GloVe & 0.242 & 0.451\\\hdashline
\convknrmrnd & \multirow{2}{*}{110,022,399} & none & 0.243 & 0.443\\
\convknrm & & GloVe & 0.268 & 0.488\\\hdashline
\bertbase & 109,483,778 & all & 0.342 & 0.585\\
\bertlarge & 335,143,938 & all & 0.353 & 0.596\\

\bottomrule
\end{tabular}
\label{tbl:models} 
\end{center}
\vspace{-0.6cm}
\end{table}
In this section, we describe the resources, models, and the process of creating the dataset, which contains a set of non-gendered queries.

\vspace{-0.1cm}
\subsection{Resources and IR Models}
\label{sec:experiments:models}
We use the MS~MARCO Passage Retrieval collection~\cite{nguyen2016ms} for our experiments. The collection consists of 8,841,822 passages, and provides a large set of informational question-style queries from Bing's search logs, accompanied by human-annotated relevant/non-relevant passages. Following Hofst{\"a}tter et al.~\cite{hofstatter2019effect}, we divide the development set of the collection into a validation and a test set. 

As classical (non-neural) IR model, we investigate BM25 with the parameters $k_1 = 0.6$ and $b = 0.8$, as suggested by Hofst{\"a}tter et al.~\cite{hofstatter2019effect}. We study the following neural ranking models: \matchpyramid~\cite{pang2016text}, Position-Aware Convolutional Recurrent Relevance Matching (\pacrr)~\cite{hui2017pacrr}, Kernel-based Neural Ranking Model (\knrm)~\cite{xiong2017end}, and Convolutional KNRM (\convknrm)~\cite{dai2018convolutional}. These models are trained using the code and the suggested parameters in Hofst{\"a}tter et al.~\cite{hofstatter2019effect}. The models re-rank the retrieval result of the BM25 model. For each model, we either initialize the word embeddings parameters randomly (denoted with $RND$ subscript), or transfer the initial values of the parameters from the 300-dimension GloVe embeddings provided by Pennington et al.~\cite{pennington2014glove}. In addition, we investigate the \bertbase and \bertlarge models, fine-tuned on the same collection by Nogueira et al.~\cite{nogueira2019passage}. The BERT models re-rank the top 200 retrieved passages of the BM25 model. Table~\ref{tbl:models} reports the number of parameters of the models. For completeness, we also report the models' evaluation results on the test set, using Mean Reciprocal Rank (MRR) and Recall at cutoff 10. The set of gender definitional words we use consists of 32 words for each gender, taken from the provided resources in previous studies~\cite{bolukbasi2016man,caliskan2017semantics}. 

In order to correctly measure gender bias in retrieval models, we need to conduct the retrieval using a set of non-gendered queries (defined in Section~\ref{sec:introduction}). In fact, if the queries contain any word or concept that refers to a specific gender, the retrieved documents are expected to be inclined towards the gender, and it does not reflect societal biases. We therefore create a dataset containing a set of non-gendered queries, explained in the next section.

\begin{table}
\begin{center}
\caption{Sample queries in the gender-annotated dataset. The number of queries is shown in parentheses.} 
\scriptsize
\vspace{-0.3cm}
\begin{tabular}{p{2.5cm} l}%
\toprule
\multirow{2}{*}{\emph{Non-gendered} (1765)$\,$} & what is a synonym for beautiful\\
& what is the meaning of resurrect\\\midrule
\multirow{2}{*}{\emph{Female} (742)} & who was oprah winfrey\\
& earliest pregnancy symptoms\\\midrule
\multirow{2}{*}{\emph{Male} (1202)} & where is martin luther king jr's place\\
& who was the king of ancient rome\\\midrule
\emph{Other or Multiple} & is g dragon gay\\
\emph{Genders} (41) & how long was shakespeare married to anne\\

\bottomrule
\end{tabular}
\label{tbl:examples} 
\end{center}
\vspace{-0.6cm}
\end{table}

\vspace{-0.3cm}
\subsection{Gender-Annotated Queries Dataset}
\vspace{-0.1cm}
Our provided gender-annotated dataset consists of a set of queries, each assigned to one of the following four categories: Non-gendered, Female, Male, Other or Multiple Genders. The Female/Male category denotes the class of queries that contain at least a word or phrase that refers to female-/male-related concepts. Such references to gender can be, for instance, gendered words like \emph{king} and \emph{queen}, words that imply gender like \emph{pregnant}, and persons or names like \emph{Nelson Mandela}. The Other or Multiple Genders category belongs to the queries that have either at least a word or phrase referring to other genders (i.e.,\ transgender and bigender), or contain references to multiple genders.

To create the dataset, we first select a set of queries of the test set that show the highest inclinations towards genders, taking into account the retrieval results of all the models. To this end, for all the IR models, we calculate the retrieval gender bias of each test set query $q$ using the \docfactortc gender magnitude measure and the $\biasrab$ approach at cutoff 10, namely $\biasrab_{10}\!(q)$ (Eq.~\ref{eq:framework:factorrab}). Using the query-level gender bias results of each model, we create two separate lists for the highest bias values towards female and male, by sorting the former from negative to positive values, and the latter reversely. Given the 7 studied IR models, we obtain 14 lists of sorted queries. We then apply the pooling method~\cite{sparck1975report} with a cutoff of 500, resulting in a total number of 3,924 unique queries. 

\begin{table}
\begin{center}
\caption{Retrieval gender bias results.}
\vspace{-0.3cm}
\scriptsize
\begin{tabular}{l l l l     l}
\toprule
\multirow{2}{*}{Model} & \multicolumn{2}{c}{\docfactortc}  & \multicolumn{2}{c}{\docfactorbool} \\\cmidrule(lr){2-3} \cmidrule(lr){4-5}
 & \multicolumn{1}{c}{RaB} & \multicolumn{1}{c}{ARaB} & \multicolumn{1}{c}{RaB} & \multicolumn{1}{c}{ARaB}  \\\midrule

\multicolumn{5}{c}{\textbf{\emph{Cut-off: 5}}} \\ 
BM25 & 0.074 & 0.074 & 0.030 & 0.025 \\\hdashline
\knrm & 0.093 ($\downarrow$0.007) & 0.088 ($\uparrow$0.010) & 0.037 ($\uparrow$0.006) & 0.036 ($\uparrow$0.010) \\
\matchpyramid & 0.093 ($\downarrow$0.014) & 0.091 ($\downarrow$0.004) & 0.050 ($\downarrow$0.013) & 0.047 ($\downarrow$0.005) \\
\pacrr & 0.102 ($\downarrow$0.017) & 0.106 ($\downarrow$0.020) & 0.041 ($\downarrow$0.002) & 0.039 ($\uparrow$0.001) \\
\convknrm & 0.076 ($\uparrow$0.002) & 0.055 ($\uparrow$0.017) & 0.039 ($\downarrow$0.004) & 0.031 ($\downarrow$0.002) \\
\bertbase & 0.106 & 0.104 & 0.059 & 0.056 \\
\bertlarge & 0.101 & 0.108 & 0.057 & 0.057 \\\hline
\multicolumn{5}{c}{\textbf{\emph{Cut-off: 10}}} \\ 
BM25 & 0.076 & 0.076 & 0.039 & 0.031 \\\hdashline
\knrm & 0.089 ($\downarrow$0.020) & 0.090 ($\downarrow$0.004) & 0.041 ($\downarrow$0.007) & 0.037 ($\uparrow$0.003) \\
\matchpyramid & 0.081 ($\downarrow$0.002) & 0.089 ($\downarrow$0.006) & 0.047 ($\downarrow$0.007) & 0.047 ($\downarrow$0.007) \\
\pacrr & 0.096 ($\downarrow$0.008) & 0.104 ($\downarrow$0.019) & 0.043 ($\uparrow$0.002) & 0.042 ($\downarrow$0.001) \\
\convknrm & 0.082 ($\downarrow$0.001) & 0.066 ($\uparrow$0.009) & 0.045 ($\downarrow$0.005) & 0.037 ($\downarrow$0.004) \\
\bertbase & 0.096 & 0.102 & 0.056 & 0.057 \\
\bertlarge & 0.090 & 0.100 & 0.052 & 0.054 \\\hline
\multicolumn{5}{c}{\textbf{\emph{Cut-off: 20}}} \\ 
BM25 & 0.074 & 0.076 & 0.041 & 0.036 \\\hdashline
\knrm & 0.083 ($\downarrow$0.010) & 0.088 ($\downarrow$0.009) & 0.044 ($\downarrow$0.005) & 0.041 ($\downarrow$0.002) \\
\matchpyramid & 0.074 ($\downarrow$0.001) & 0.083 ($\downarrow$0.004) & 0.043 ($\downarrow$0.004) & 0.046 ($\downarrow$0.006) \\
\pacrr & 0.086 ($\downarrow$0.009) & 0.098 ($\downarrow$0.013) & 0.043 ($\downarrow$0.001) & 0.043 ( 0.000) \\
\convknrm & 0.084 ($\downarrow$0.005) & 0.075 ($\uparrow$0.003) & 0.049 ($\downarrow$0.005) & 0.042 ($\downarrow$0.005) \\
\bertbase & 0.097 & 0.099 & 0.057 & 0.056 \\
\bertlarge & 0.082 & 0.093 & 0.053 & 0.053 \\\hline
\multicolumn{5}{c}{\textbf{\emph{Cut-off: 30}}} \\ 
BM25 & 0.073 & 0.075 & 0.041 & 0.038 \\\hdashline
\knrm & 0.074 ($\downarrow$0.003) & 0.086 ($\downarrow$0.008) & 0.041 ($\downarrow$0.002) & 0.042 ($\downarrow$0.003) \\
\matchpyramid & 0.070 ($\uparrow$0.001) & 0.080 ($\downarrow$0.003) & 0.040 ($\downarrow$0.001) & 0.044 ($\downarrow$0.005) \\
\pacrr & 0.083 ($\downarrow$0.010) & 0.094 ($\downarrow$0.012) & 0.045 ($\downarrow$0.005) & 0.043 ($\downarrow$0.001) \\
\convknrm & 0.082 ($\downarrow$0.008) & 0.078 ( 0.000) & 0.049 ($\downarrow$0.007) & 0.045 ($\downarrow$0.005) \\
\bertbase & 0.092 & 0.097 & 0.055 & 0.056 \\
\bertlarge & 0.083 & 0.090 & 0.052 & 0.053 \\

\bottomrule
\end{tabular}
\label{tbl:results} 
\end{center}
\vspace{-0.7cm}
\end{table}

\begin{figure*}[tb]
\centering
\includegraphics[width=0.95\textwidth]{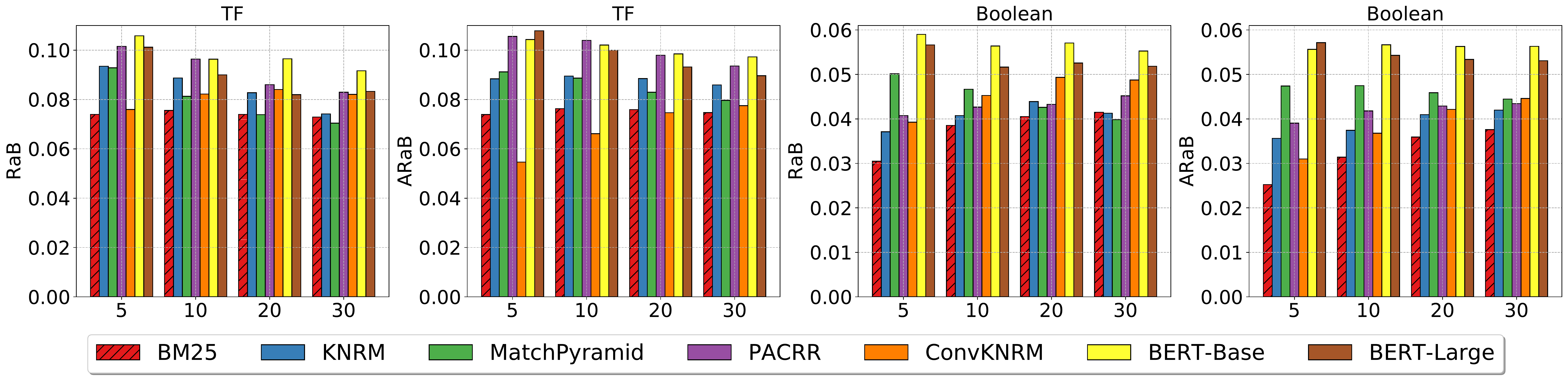}
\vspace{-0.3cm}
\caption{Results of retrieval gender bias metrics. Higher (positive) values indicate higher bias towards male in retrieval results.}
\label{fig:results}
\vspace{-0.5cm}
\end{figure*}

Using these queries, in the next step, we ask three Amazon Mechanical Turk workers to classify each query to one of the four categories. The details and descriptions of the annotation experiments are available in the provided resources.\footnote{Dataset, resources, and code are available on \emph{\url{https://github.com/navid-rekabsaz/GenderBias_IR}}.} Based on the annotation results, we assign a category to each query using the majority vote of the annotators. We remove the queries for which no unambiguous majority decision could be made (i.e.\, all three annotators selected a different category). This results in the final dataset of 3,750 queries. Table~\ref{tbl:examples} shows some samples of the dataset and the overall number of queries in each category. In our subsequent experiments, we only use the 1,765 non-gendered queries.

%% file: 4-results.tex
The retrieval gender bias results of the models, measured with the \biasrab and \biasArab metrics on the cutoffs of 5, 10, 20, and 30, for both document gender magnitude measures \docfactortc and \docfactorbool, are shown in Table~\ref{tbl:results}. In the results of the neural ranking models (except the BERT-based ones), the reported values belong to the models with pre-trained GloVe embeddings; the values in parentheses indicate the changes in the retrieval gender bias values in comparison to the ones of the corresponding $RND$ models (when the word embeddings are initialized randomly). An arrow down/up in the parentheses indicates a decrease/increase in the bias when using a $RND$ model. For easier visual comparison, Figure~\ref{fig:results} depicts the results in plots.

The results show the inclination of all the IR models towards the male concepts (despite using non-gendered queries). The neural models consistently increase retrieval gender bias in comparison with BM25 in almost all variations (4 exceptions out of 96 variations). The BERT models, and especially \bertbase, show the overall highest degrees of gender bias. These confirm that the neural models, despite better retrieval performance, on the whole intensify gender bias in retrieval results toward male when compared with BM25. 

Finally, we look at the effect of using pre-trained word embeddings on the retrieval gender bias of neural ranking models. Based on the results, transfer learning either increases (cases with down arrows) or does not affect (0 values) gender bias in the majority of the cases, namely 53 out of 64 cases. This therefore shows that transfer learning tends to increase gender bias in retrieval results.